\documentclass[11pt]{article}
\input{epsf}
\usepackage{graphicx}
\usepackage{subcaption}
\usepackage{multirow}
\usepackage{comment}
\usepackage{indentfirst}
\usepackage{amssymb}

\let\oldthebibliography\thebibliography
\renewcommand\thebibliography[1]{
  \oldthebibliography{#1}
  \setlength{\itemsep}{5pt}
}







\begin{document}
\addtolength{\baselineskip}{0.01in}

\vspace*{0.5cm}

\begin{center}
{\large\bf
A Review of Proximity Effect Correction in Electron-beam Lithography
}
\end{center}

\vspace{1mm}
\begin{center}
Pengcheng Li\\
Department of Electrical and Computer Engineering \\
Auburn University, Auburn, AL, USA \\
\end{center}

\bigskip

\begin{center}
{\bf Abstract}
\end{center}
I review the work of proximity effect correction (PEC) in electron-beam (e-beam) lithography with emphasis on dose modification and shape modification PEC techniques.

\section{Proximity Effect Correction}

Electron-beam (e-beam) lithography is a lithographic process used to transfer circuit patterns onto silicon or other substrates. It employs a focused beam of electrons to expose a circuit pattern into the electron-sensitive resist applied to the surface of the substrate. The main advantage of e-beam lithography is that it can offer much higher patterning resolution than the conventional optical lithography which is limited by the diffraction of light, thus it is extremely suitable for fabrication of nanoscale features.

The proximity effect in e-beam lithography is an effect due to forward-scattering of electrons in the resist and backscattering of electrons from the substrate. The electron scattering leads to undesired exposure (e-beam energy deposited) of the resist in the unexposed regions adjacent to those exposed by the e-beam, which in turn causes changes in the dissolution rate of the resist. Therefore, the unexposed regions receiving the scattered electrons are also partially developed, which results in circuit patterns with dimensions different from the target ones.

In general, the scattering of electrons causing the proximity effect can be modeled as a convolution of the dose (e-beam energy given) distribution of a circuit pattern with a proximity function \cite{Chang1975}-\cite{Demers2011}. The proximity function, usually referred to as the point spread function (PSF), is radially symmetric and shows how the electron energy is distributed throughout the resist when a single point is exposed.

It is well known that the proximity effect can be reduced or corrected by appropriate measures, including physical techniques and software-based techniques \cite{Parikh1978}-\cite{Eichfeld2014}.

The physical techniques are to modify the physical conditions of the e-beam lithography systems or the physical characteristics of resists, such as high beam energy technique \cite{Kyser1979}, low beam energy technique \cite{Kyser1979,McCord1992,Stark1993}, substrate material optimization technique \cite{Aizaki1979}, multilayer resists technique \cite{Kruger1981}, intermediate layer technique \cite{Dobisz1993}, etc.

Experimental results show that these physical techniques provide a possible way to reduce the proximity effect, however they can only reduce it to some extent, with drawbacks and limitations. For example, the multilayer resists technique requires a complicated resist process step which introduces extra complexity and does not work for circuit patterns with feature size of 1 $\mu$m or less. The high beam energy technique and the low beam energy technique are also affected by the backscattering range and the resist thickness, respectively.

Therefore, most techniques was focused on developing practical and effective software-based proximity effect correction (PEC) schemes in order to eliminate the proximity effect completely, which is presented in detail as below.

\section{Dose Modification PEC Techniques}

\subsection{Self-consistent Method}

The first work in this specific type of PEC schemes was done by Parikh in 1978 \cite{Parikh1978,Parikh1979_1,Parikh1979_2,Parikh1979_3}, who developed a self-consistent PEC method for resist exposure. The purpose of this PEC method is to compute the dose which must be applied in order to obtain identical average absorbed (incident plus backscattered) exposure in the resist in each written shape of the pattern addressed by the e-beam.

The self-consistent PEC method can be formulated mathematically by solving the following set of linear equations:

\begin{equation}
\left\{
\begin{array}{cl}
E_T = I_{11} \cdot D_1 + I_{12} \cdot D_2 + \cdots I_{1N} \cdot D_N\\
E_T = I_{21} \cdot D_1 + I_{22} \cdot D_2 + \cdots I_{2N} \cdot D_N\\
\vdots \\
E_T = I_{N1} \cdot D_1 + I_{N2} \cdot D_2 + \cdots I_{NN} \cdot D_N\\
\end{array},
\right.
\label{eqn:Parikh_1}
\end{equation}
where $N$ is the number of shapes, $E_T$ is the required target exposure for each shape, $D_i$ is the dose given for shape $i$, and $I_{ij}$ is the proximity interaction between shape $i$ and shape $j$.

The integral defining the proximity interaction $I_{ij}$ between shape $i$ and shape $j$ with area $A_i$ and area $A_j$ is given by:

\begin{equation}
I_{ij} = \int_{A_i}\int_{A_j}\mbox{psf}(r)dA_jdA_i,
\label{eqn:Parikh_2}
\end{equation}
where $\mbox{psf}(r)$ is the PSF.

In general, the integral in Eq. \ref{eqn:Parikh_2} cannot be evaluated explicitly for two arbitrary shapes, i.e., a general formula does not exist. However, Parikh derived an analytical expression for evaluating the proximity interaction between two rectangular shapes. For any other types of shape, such as trapezoid, it can be approximated by a set of rectangles, which has the same total area as the original shape. Then the proximity interaction between two arbitrary shapes can be approximated by a summation of the proximity interactions between these rectangular components. Since the proximity interaction between two rectangular shapes can be computed rather exactly, one can split the whole pattern into rectangles before correction and set up the proximity interaction matrix $I$ in Eq. \ref{eqn:Parikh_1}. Once the proximity interaction matrix $I$ is computed, the $N$ linear equations in Eq. \ref{eqn:Parikh_1} can be uniquely solved for each $D_i$, i.e., the corrected dose for each shape.

Later many researchers developed various PEC methods based on the self-consistent PEC method to further improve its correction accuracy and efficiency.

One limitation of the self-consistent PEC method is that it relies only on the constraint that the average exposure in the resist is identical within the exposed area of each written shape. Therefore, it does not take the exposure in the regions not addressed by the e-beam into account. In order to solve this issue, Parikh developed another PEC method, i.e., the unaddressed-region compensation method \cite{Parikh1979_1}, which attempts to account for the exposure in the unaddressed regions between shapes. The unaddressed regions, also known as the unexposed area, receive undesired exposure due to electron scattering, which has to be minimized. Therefore, the purpose of this PEC method is to make the exposure below a specific value in the unexposed area while the identical average exposure is still received in each exposed shape. Note that this PEC method can be considered as an expansion of the self-consistent PEC method as the latter one is a special case of the former one when the exposure threshold for the unexposed area becomes large enough.

In Parikh's self-consistent PEC method, average exposure in each exposed shape is computed by integrating the exposure over the shape. However, Phang and Ahmed pointed out that the exposure value at the edge of the exposed shape is more important for obtaining accurate pattern fidelity of the fabricated structure. Based on this conclusion, they developed a PEC method for line patterns \cite{Phang1979}. In this PEC method, for an exposed line element, a dose is required such that the edge of the exposed line element has an exposure equal to $E_T$, i.e., the threshold of exposure for the dissolution of the resist. This is the fundamental principle involved in obtaining pattern fidelity between the designed and the actual fabricated structure. Similar to the self-consistent PEC method, an exposure equation to compute the the exposure at the edge of a line element is set up for each exposed line element. Then the dose of each line element can be derived by solving a set of linear equations.

Carroll pointed out that Parikh's self-consistent PEC method is not general as its model is based on equality constraints which requires exactly as many constraints as variables. He developed a PEC method based on inequality constraints which is far more general and can have arbitrarily many constraints \cite{Carroll1981_1, Carroll1981_2}. In this PEC method, $N$ pixels on the resist surface is chosen for exposure computation. For those pixels in the exposed area, their exposure values are required to be not smaller than a target threshold $E_{T1}$ (but not have to be equalized), while for those pixels in the unexposed area, their exposure values are required to be not larger than another target threshold $E_{T2}$. Through this way, $N$ inequality constraints can be formed with $N$ dose values as variables. In general, a system of this form is usually underdetermined, thus a cost function is employed to help solving this underdetermined system. In the cost function, the total dose which represents the total exposure time, is to be minimized. The best solution can be derived through linear programming approach.

As mentioned before, the computation of average exposure results in a single exposure value for each shape, i.e., each exposed shape is considered as a whole in Parikh's self-consistent PEC method. Kratschmer developed a new PEC method in which each exposed shape is partitioned into elements in order to gain more flexibility in dose control and thus achieve better exposure distribution \cite{Kratschmer1981}. In this PEC method, each shape is first partitioned to account for the intra proximity effect, then the subshapes at the edges of each shape are further partitioned to account for the inter proximity effect. The setting up and solving of exposure equations are the same as in the self-consistent PEC method, but here each exposure equation corresponds to a partitioned element rather than a whole shape. By appropriate partitioning, the number of linear equations can be limited to reduce the computation time.

For a pattern of $N$ shapes, the implementation of the self-consistent PEC method requires the computation of $N^2$ proximity interaction coefficients $I_{ij}$. This number can be reduced to $N(N+1)/2$ as matrix $I$ is symmetric ($I_{ij} = I_{ji}$) according to the reciprocity principle, which however is still a significant amount of computation. A series of approaches was employed in order to reduce the computation time, including Otto and Griffith's parallel processing approach \cite{Otto1988_1,Otto1988_2} and Vermeulen et al.'s clustering approach \cite{Vermeulen1989}. By applying those approaches, the linear equation set in Eq. \ref{eqn:Parikh_1} is decomposed into a number of independent linear equation sets of lower order which are handled independently, thus the correction speed can be remarkably increased.

\subsection{Transform-based Method}

In 1983, Chow et al. proposed that the PEC problem can be solved by employing conventional image processing approaches, based on which they developed a transform-based PEC method \cite{Chow1983}.

In this PEC method, the required dose distribution $d(x)$ is solved through direct deconvolution of the target exposure distribution $e(x)$ with the PSF $\mbox{psf}(x)$, which is given by:

\begin{equation}
d(x) = e(x) \circledast^{-1} \mbox{psf}(x).
\label{eqn:Chow_1}
\end{equation}

The above deconvolution can also be implemented in the frequency domain using the forward and inverse Fourier transforms, which is given by:

\begin{equation}
d(x) = F^{-1}\left[\frac{F[e(x)]}{F[\mbox{psf}(x)]}\right].
\label{eqn:Chow_2}
\end{equation}

Note that unlike the above-mentioned self-consistent PEC methods in which the corrected dose is constant within each shape or partitioned element, the transform-based PEC method is based on the pixel level, thus derives an extremely accurate solution for the dose distribution which is almost exact.

However, this simple PEC method suffers from two major problems. The first problem is that the solved dose distribution has negative values at some locations due to the Gibbs phenomenon from the Fourier transform, which cannot be physically realized. To solve this problem, the method applies corner rounding to the target exposure distribution to minimize the negative dose values in the solution. Then the remaining negative values is further removed by adding a constant to the solved dose distribution to sufficiently guarantee the positive condition. The second problem is that the solved dose distribution contains too much details from rapid oscillations, thus results in an extremely large data base, which is unmanageable and inefficient. The method solve this problem based on an approximation process by introducing the Walsh transform thinning algorithm to reduce the detail part while preventing the deterioration of the effective dose. Through Walsh transform, the data base compression is achieved, and the final data is stored in the form of its Walsh coefficients. Since the Fourier transform can be implemented using its fast algorithm on special purpose computing hardware, the computation time of this method can be greatly reduced.

Haslam et al. later further improved this PEC method and verified it through experiments \cite{Haslam1985,Haslam1986,Haslam1988}. The most notable improvement was to extent the application of this method from simple one-dimensional line patterns to two-dimensional general patterns. In order to achieve further data compression, the data thinning algorithm employs the more powerful two-dimensional Haar transform instead of the previous Walsh transform. An additional benefit of using the Haar transform is that its fast algorithm can be implemented by the same computing hardware which is used for the fast Fourier transform. Eisenmann et al. later achieved a further reduction of the computation time by separating the calculation into correction related and pattern related steps \cite{Eisenmann1993,Waas1995}.

\subsection{Approximate Formula-based Method}

In 1989, Abe et al. found that the forward-scattered electron range can be assumed to be negligibly small especially for the cases of high beam energy or thin resist thickness. Based on this assumption, they developed a PEC method which utilizes an approximate dose correction formula \cite{Abe1989_2}.

In this PEC method, the corrected dose $D(x)$ for a specific feature is derived by the following approximate formula:

\begin{equation}
D(x) = D_0(1 - kU(x)),
\label{eqn:Abe_1}
\end{equation}
where $D_0$ is the initial dose for each feature, and $k$ is a parameter selected to minimize the relative error of exposure. The function $U(x)$ corresponds to the exposure caused by the backscattered electrons, which is given by:

\begin{equation}
U(x) = \frac{1}{\pi\beta^2} \int_A \exp\left[-\frac{(x - x')^2}{\beta^2}\right]dx',
\label{eqn:Abe_2}
\end{equation}
where $\beta$ is as described in the double-Gaussian PSF (refer to Eq. \ref{eqn:Chang}), and the region $A$ is the exposed area within $3\beta$ from the center of the feature.

The first and second terms of Eq. \ref{eqn:Abe_1} corresponds to the initial dose and the background dose by the backscattered electrons, respectively. Therefore, the corrected dose $D(x)$ for a specific feature is approximated by subtracting its background dose from the initial dose $D_0$. In order to reduce the computation time, the function $U(x)$ can be modified into another form which contains only the mathematical error function $\mbox{erf}(z)$ by dividing or approximating all the features of the circuit pattern into rectangles, where the values of the error function table can be predetermined for fast referring.

Abe et al. later further accelerate the correction speed by adding the pattern inversion method and the data compaction method, and installed it into a high beam energy e-beam direct writing system \cite{Abe1989_1}. A more powerful representative figure method is later employed to further reduce the computation time with negligible error being introduced \cite{Abe1991,Abe1996}.

\subsection{Pattern Area Density Map Method}

In 1992, Murai et al. developed a PEC method adjusting dose by referring to a pattern area density map which is calculated by utilizing the fixed size of square meshes \cite{Murai1992}.

In this PEC method, the circuit pattern is partitioned with a fixed sized mesh. The mesh size is chosen such that the variation of exposure by the backscattered electrons within a single mesh site is negligible. The pattern area density $\lambda$, which is defined as the ratio of the exposed area to the total area in a region, is computed in each mesh site, giving a pattern area density map for the circuit pattern. The $\lambda$ map is then convolved with a smoothing filter, giving a smoothed $\lambda$ map, i.e., $\lambda_{sm}$ map. Each circuit feature is then partitioned into rectangles, and a $\lambda'$ value is assigned to each rectangle which is the linear interpolation of the $\lambda_{sm}$ values of the four nearest mesh sites to the rectangle. Finally, a corrected dose is assigned to each rectangle which is given by:

\begin{equation}
D = C \frac{2(1 + \eta)}{1 + 2\lambda'\eta},
\label{eqn:Murai}
\end{equation}
where $\eta$ is as described in the double-Gaussian PSF (refer to Eq. \ref{eqn:Chang}), and $C$ is a constant depending on the specific resist and beam energy used.

The determination of the mesh size is important because the correction error depends on the mesh size and the smoothing range. In order to reduce the error, a small mesh size and a wide smoothing range are desirable, which however needs to be compromised with the amount of map size. The smoothing filter is based on a form of template convolution where the forward scattering is neglected for fast computation purpose.

Kasuga et al. later improved the correction accuracy of this PEC method by employing an adaptive partition and dose adjustment algorithm based on the gradient vector from the pattern area density map \cite{Kasuga1996}. Ea and Brown further enhanced this PEC method by reducing the correction error using an iterative algorithm and a framing procedure \cite{Ea1999}, and incorporating it with a corner correction scheme \cite{Ea2001}.

\subsection{Other Methods}

Other dose modification PEC methods include Greeneich's dose compensation curve method \cite{Greeneich1981}, Pavkovich's integral equation approximate solution method \cite{Pavkovich1986}, Gerber's splitting equation exact solution method \cite{Gerber1988}, Frye's adaptive neural network method \cite{Frye1991}, Aristov et al.'s simple compensation method \cite{Aristov1992}, Rau et al.'s nonlinear optimization method \cite{Rau1996}, Watson et al.'s inherent forward scattering correction method \cite{Watson1997}, etc.

\section{Shape Modification PEC Techniques}

\subsection{Empirically-determined Method}

The history of the shape modification PEC techniques began as early as the dose modification PEC techniques. In 1978, Sewell developed the first PEC method in this specific area \cite{Sewell1978}. The core part of this PEC method is that although the change in the pattern dimensions during fabrication depends on parameters such as beam current, beam scan frequency, developing time, developing temperature, etc., there is a basic relationship between the designed pattern dimensions and the pattern dimensions after development. Therefore, the changes in the designed pattern dimensions to compensate the influence of proximity effect could yield patterns with the correct dimensions in the resist. By analyzing the changes in the actual measured pattern dimensions for different amount of dose values from experimental results, a set of empirical curves is derived to set up design tables. For a design table, a specific feature is chosen as the control feature to set the exposure condition such that all resist exposed above a specific value is developed, and the rest of the pattern dimensions are all adjusted using this exposure threshold. Based on the design table, the pattern dimensions after development can be predicted, and the designed pattern dimensions are adjusted to compensate the proximity effect.

\subsection{Analytically-determined Method}

In 1979, Parikh analytically modeled the relationship between the designed pattern dimensions and the pattern dimensions after development, and pointed out that this leads to an underdetermined system of nonlinear equations, whose solution is extremely difficult and impractical for an arbitrary shape \cite{Parikh1979_1}. However, for simple shapes such as squares and infinitely long lines, the solution is possible to derive as only one dimension variable (e.g., width) needs to be considered for each shape. In 1980, he developed a shape-dimension adjustment method in which exact solutions can be obtained for simple shapes based on analytical calculation instead of the empirical method \cite{Parikh1980}.

In this PEC method, the formulation of intra proximity effect and inter proximity effect lead to a set of simultaneous nonlinear equations, which is individually given by.

\begin{equation}
E_T = \frac{1}{2(1 + \eta)}\left[f(\alpha) + \eta f(\beta)\right],
\label{eqn:Parikh_3}
\end{equation}
where $\alpha$, $\beta$, and $\eta$ are as described in the double-Gaussian PSF (refer to Eq. \ref{eqn:Chang}), $E_T$ is the threshold of exposure for the dissolution of the resist, and the function $f(x)$ is given by:

\begin{equation}
f(x) = \mbox{erf}\left(\frac{w_a}{2x}\right)\left[\mbox{erf}\left(\frac{w_a - w_d}{2x}\right) + \mbox{erf}\left(\frac{w_a + w_d}{2x}\right)\right],
\label{eqn:Parikh_4}
\end{equation}
where $w_d$ is the designed width of the pattern while $w_a$ is the actual width of the pattern after development, and $\mbox{erf}(z)$ is the error function as defined in mathematics.

For the case of isolated features which consider only intra proximity effect, only one nonlinear equation for the exposure at the edge of a square or line is required to solve for the designed width of the square or line, i.e., $w_d$. For the case of interacting features which consider both intra proximity effect and inter proximity effect, a set of nonlinear equations for the exposure at different locations are required which are much more difficult in solving. For a simple case such as an infinitely long line adjacent to a large area, both the exposure at the left and right edge of the line are calculated to set up nonlinear equations to solve for the designed width of the line, i.e., $w_d$.

\section{Distinctive PEC Techniques}

\subsection{Background Exposure Equalization Method}

In 1983, Owen and Rissman developed a PEC method, GHOST, which does not suffer from the lengthy and costly computation required in the dose and shape modifications \cite{Owen1983}. The correction scheme of GHOST is based on equalization of the background exposure, i.e., backscattered electron dose, received by all points within a pattern.

The pattern is first exposed (to generate a pattern exposure) using a focused beam with a dose $D_e$. Then the reverse field of the pattern is exposed (to generate a correction exposure) using a defocused electron beam with a diameter $d_c$:

\begin{equation}
d_c = \frac{\beta}{(1 + \eta)^{1/4}},
\label{eqn:Owen_1}
\end{equation}

and a reduced dose $D_c$:

\begin{equation}
D_c = D_e \cdot \frac{\eta}{1 + \eta},
\label{eqn:Owen_2}
\end{equation}
where $\beta$ and $\eta$ are as described in the double-Gaussian PSF (refer to Eq. \ref{eqn:Chang}).

By making a correction exposure in addition to the pattern exposure for equalizing the background exposure, the compensation for the proximity effect is achieved. As a result, all the features in the pattern will develop out in a more uniform manner in the resist development stage. Note that the implementation of GHOST only requires image reversal of the pattern data without any further computation.

Since the correction scheme of GHOST is simple and general enough to be implemented on any type of e-beam lithography systems, its application was further extended with various patterns and e-beam parameter settings \cite{Owen1985,Kostelak1988_1,Kostelak1988_2,Leen1989,Muray1990,Owen1990,Liu1990,Moriizumi1993}. Watson et al. later improved GHOST by applying the pattern exposure and the correction exposure at the same time to eliminate the throughput drawback of the direct-write GHOST \cite{Watson1995,Watson1998}.

\subsection{Hierarchical Rule-based Method}

In 1991, Lee et al. developed a PEC method, PYRAMID, which has a hierarchical rule-based correction scheme \cite{Lee1991,Lee1998_2,Cook1998}. One of the most distinct features of PYRAMID is its two-level hierarchy in exposure estimation and correction, which is not restricted to either dose modification or shape modification. The first version and early improvements of PYRAMID adopted a shape modification for the correction part. However, there is nothing inherent in PYRAMID that requires shape modification, later the same overall correction hierarchy was extended to dose modification and hybrid (dose/shape) modification \cite{Cook1997,Lee2001_1,Hu2003,Lee2003}.

Based on the digital image processing model of e-beam lithography, the exposure estimation of PYRAMID is implemented by calculating the total exposure as the sum of two separated components, global exposure and local exposure. The reason that the exposure can be separated into two components stems directly from the fact that the PSF can be considered as consisting of the sum of two distinct components, i.e., a sharp, short range local component and a flat, long range global component. The global exposure is an approximation of the exposure due to circuit features located far away from the critical point and is calculated through a coarse grain convolution. The process of calculating global exposure begins by producing a coarse image of the circuit pattern by dividing it into large pixels or global exposure blocks, where the value of each block is the circuit area contained within the corresponding global exposure block in the circuit pattern. The global exposure can then be found by convolving the coarse image with a 2-D sampled version of the PSF, which has a sampling pixel size equal to the global exposure block size. The local exposure, as opposed to the global exposure, is exact and considers circuit area located close to the critical point. The local exposure at a given point is calculated by applying exact convolution to all circuit area falling within a small window centered about the global exposure block containing the point. This window is termed the local exposure window and serves to separate circuit area contributing to global exposure (area outside the window) from area contributing to local exposure (area inside the window).

Similarly, the correction procedure of PYRAMID is also divided into two parts, local correction and global correction. The local correction attempts to adjust the local dose distribution (for dose modification) or the pattern shape (for shape modification) to compensate for intra proximity effect and inter proximity effect caused by all the features within a small window. The local correction itself is very systematic, using two levels of correction to minimize intra proximity effect and inter proximity effect, respectively. Rule tables are used to dictate correction modes for different situations. Special consideration is given to correction of features which are in close proximity or connected at junctions. While the local correction ignores interactions between widely separated features, the global correction takes general characteristics of the entire circuit pattern into account to makes adjustments to the local corrections based on differences in exposure values at various geometric locations in the circuit pattern.

\section{Hybrid PEC Techniques}

In order to achieve better correction accuracy and efficiency, several hybrid PEC methods have been proposed. In a hybrid PEC method, two or more independent PEC methods are usually contained, e.g., a combination of dose modification method and shape modification method.

In Groves's hybrid PEC method, the analytically determined shape modification method is adopted to offset forward scattering while the self-consistent dose modification method is adopted to offset backscattering \cite{Groves1993}.

Later, Cook and Lee developed a hybrid PEC method for PYRAMID \cite{Cook1997}. In this hybrid PEC method, a circuit feature is partitioned into regions for region-wise dose control, where each region is assigned a different dose using the self-consistent dose modification method and the hierarchical rule-based shape modification method is carried out within each region.

Wind et al. developed a suite of PEC programs incorporating several dose modification methods, where the self-consistent dose modification method is used for forward-scattering correction and the pattern area density map dose modification method is used for backscattering correction \cite{Wind1998}.

Takahashi et al. separated the forward-scattering correction and the backscattering correction such that the analytically-determined shape modification method is used for the forward-scattering and is performed only once for repeated features while the pattern area density map dose modification method is used for backscattering correction and is iterated for reducing error \cite{Takahashi2000,Osawa2001,Ogino2003}.

Recently, Klimpel et al. developed a model-based hybrid PEC method, where the PSF (the proximity function) is modeled in different forms for the forward-scattering correction and the backscattering correction, respectively \cite{Klimpel2011}. Based on this model, an iterative shape modification method is adopted for forward-scattering correction while an iterative dose modification method is adopted for backscattering correction.

\section{Summary}

In sum, most existing PEC techniques can be classified into two types: (1) Adjustment of incident electron dose, i.e., dose modification. This is achieved by appropriate variations in the dwell time or the beam intensity of a e-beam scanning system. (2) Adjustment of pattern dimensions, i.e., shape modification. This is achieved by appropriate changes in the pattern dimensions such that patterns with the desired dimensions are obtained after exposure and development.

The dose modification PEC technique has the advantage that it provides a mathematically unique solution which depends only on the form and the magnitude of the PSF. This PEC technique is general enough to apply to an arbitrary definition of a region, from the smallest electron beam defined element to an entire shape. However, this PEC technique requires a lengthy and costly computation to evaluate the correction and a large database to store the dose value of each region. Furthermore, it cannot be applied on some certain kinds of e-beam lithography systems, such as the e-beam projection systems, which do not have the freedom of being able to change the dose for each shape as the pattern is being exposed.

The shape modification PEC technique has the advantage that it does not require sophisticated dose adjustment, thus is much more simple than the dose modification PEC technique. Also in the case of e-beam projection systems, the shape modification PEC technique is a better choice as it requires a relatively simple pattern generator in terms of not changing the beam frequency and a relatively simple computer programs for correcting pattern dimensions. The main disadvantage of this PEC technique is similar to the dose modification PEC technique, i.e., requiring a lengthy and costly computation to evaluate the correction. Its correction accuracy is lower than the dose modification PEC technique. Furthermore, the minimum magnitude of the shape changes is limited by the resolution of e-beam lithography systems.

Other PEC techniques, such as the GHOST method, has the advantage that it is simple and general enough to be implemented on any type of e-beam lithography systems and does not suffer from the time-consuming computation and large database storage required in the dose and shape modifications. The main disadvantage of this PEC technique is that the throughput of the lithographic process is reduced, typically by a factor of 2, due to compensating the proximity effect by a correction exposure. Another drawback is the contrast degradation between the pattern exposure and the background exposure which leads to a dull resist profile.

\end{document}